\documentclass[twocolumn,twocolappendix]{aastex631}
\usepackage{multirow}
\usepackage{color}
\usepackage{graphicx}
\usepackage{subfigure}
\usepackage{natbib}
\usepackage{booktabs}
\usepackage{chngcntr}

\usepackage[encapsulated]{CJK}
\usepackage{mathrsfs}
\usepackage{amsmath}
\usepackage{epstopdf}
\usepackage{longtable}
\usepackage{booktabs}
\usepackage{hyperref} 

\newbox\grsign \setbox\grsign=\hbox{$>$} \newdimen\grdimen \grdimen=\ht\grsign
\newbox\laxbox \newbox\gaxbox
\setbox\gaxbox=\hbox{\raise.5ex\hbox{$>$}\llap
     {\lower.5ex\hbox{$\sim$}}}\ht1=\grdimen\dp1=0pt
\setbox\laxbox=\hbox{\raise.5ex\hbox{$<$}\llap
     {\lower.5ex\hbox{$\sim$}}}\ht2=\grdimen\dp2=0pt

\date{\today}
\submitjournal{ApJL}
\shorttitle{Spiral Structure}
\shortauthors{Sun et al.}

\newcommand{\s}{\,{\rm s}}

        \newcommand{\K}{\,{\rm K}}
\newcommand{\kms}{km\,s$^{\rm -1}$}

\newcommand{\vlsr}{$v_{\rm lsr}$}
\graphicspath{{./}{figures/}}
\begin{document}

\title{A new view of the Spiral Structure of the Northern Outer Milky Way in Carbon Monoxide}
\correspondingauthor{Yan Sun \& Ji Yang}
\email{yansun@pmo.ac.cn, jiyang@pmo.ac.cn}

\author[0000-0002-3904-1622]{Yan Sun}
\affiliation{Purple Mountain Observatory, Chinese Academy of Sciences, Nanjing 210008, China}

\author[0000-0001-7768-7320]{Ji Yang}
\affiliation{Purple Mountain Observatory, Chinese Academy of Sciences, Nanjing 210008, China}

\author[0000-0003-2549-7247]{Shaobo Zhang}
\affiliation{Purple Mountain Observatory, Chinese Academy of Sciences, Nanjing 210008, China}

\author[0000-0003-4586-7751]{Qing-Zeng Yan}
\affiliation{Purple Mountain Observatory, Chinese Academy of Sciences, Nanjing 210008, China}
\author[0000-0002-0197-470X]{Yang Su}
\affiliation{Purple Mountain Observatory, Chinese Academy of Sciences, Nanjing 210008, China}
\author[0000-0003-3151-8964]{Xuepeng Chen}
\affiliation{Purple Mountain Observatory, Chinese Academy of Sciences, Nanjing 210008, China}
\author[0000-0003-2418-3350]{Xin Zhou}
\affiliation{Purple Mountain Observatory, Chinese Academy of Sciences, Nanjing 210008, China}
\author[0000-0001-5602-3306]{Ye Xu}
\affiliation{Purple Mountain Observatory, Chinese Academy of Sciences, Nanjing 210008, China}
\author[0000-0003-0746-7968]{Hongchi Wang}
\affiliation{Purple Mountain Observatory, Chinese Academy of Sciences, Nanjing 210008, China}
\author{Min Wang}
\affiliation{Purple Mountain Observatory, Chinese Academy of Sciences, Nanjing 210008, China}
\author{Zhibo Jiang}
\affiliation{Purple Mountain Observatory, Chinese Academy of Sciences, Nanjing 210008, China}
\author{Ji-Xian Sun}
\affiliation{Purple Mountain Observatory, Chinese Academy of Sciences, Nanjing 210008, China}
\author{Deng-Rong Lu}
\affiliation{Purple Mountain Observatory, Chinese Academy of Sciences, Nanjing 210008, China}
\author{Bing-Gang Ju}
\affiliation{Purple Mountain Observatory, Chinese Academy of Sciences, Nanjing 210008, China}
\author{Xu-Guo Zhang}
\affiliation{Purple Mountain Observatory, Chinese Academy of Sciences, Nanjing 210008, China}
\author{Min Wang}
\affiliation{Purple Mountain Observatory, Chinese Academy of Sciences, Nanjing 210008, China}

\begin{abstract}

Based on 32,162 molecular clouds from the Milky Way Imaging Scroll Painting project, we obtain new face-on molecular gas maps of the northern outer Galaxy. The total molecular gas surface density map reveals three segments of spirals, extending 16–43 kiloparsecs in length. The Perseus and Outer arms stand out prominently, appearing as quasi-continuous structures along most of their length. At the Galactic outskirts, about 1,306 clouds 
connect the two segments of the new spiral arm discovered by \cite{2011ApJ...734L..24D} in the first quadrant and \cite{2015ApJ...798L..27S} in the second quadrant, possibly extending the arm into the outer third quadrant. Logarithmic spirals can be fitted to the CO arm segments with pitch angles ranging from 4$^{\circ}$ to 12$^{\circ}$. 
These CO arms extend beyond previous CO studies and the optical radius, reaching a galactic radius of about 22 kiloparsecs, comparable to the H{\sc i} radial range. 
\end{abstract}


\section{Introduction} 
Spiral galaxies have long captivated astronomers with their striking spiral structures, characterized by active star formation and dense interstellar medium (ISM). Among them, the Milky Way has been proposed as a spiral galaxy for centuries~\citep{1852AJ......2..158A}, but its observed spiral patterns remain uncertain~\citep[e.g.,][]{2009IAUS..254..319B,2014PASA...31...35D,2018RAA....18..146X}. The challenge primarily stems from our position deep within the galactic disk, 
where the presence of abundant interstellar dust renders optical studies ineffective beyond several kiloparsecs~\citep[e.g.,][]{2006Sci...312.1773L}.
Radio wavelength observations, particularly the H{\sc i} and CO line emissions, offer a promising avenue for exploring of the gaseous disk extending far beyond the stellar disk~\citep[e.g.,][]{1969ARA&A...7...39K,2004Natur.432..369B,2009ARA&A..47...27K,2015ARA&A..53..583H}. 
\begin{figure*}
\vspace{-1mm}
\centering
\includegraphics[angle=0,scale=0.35]{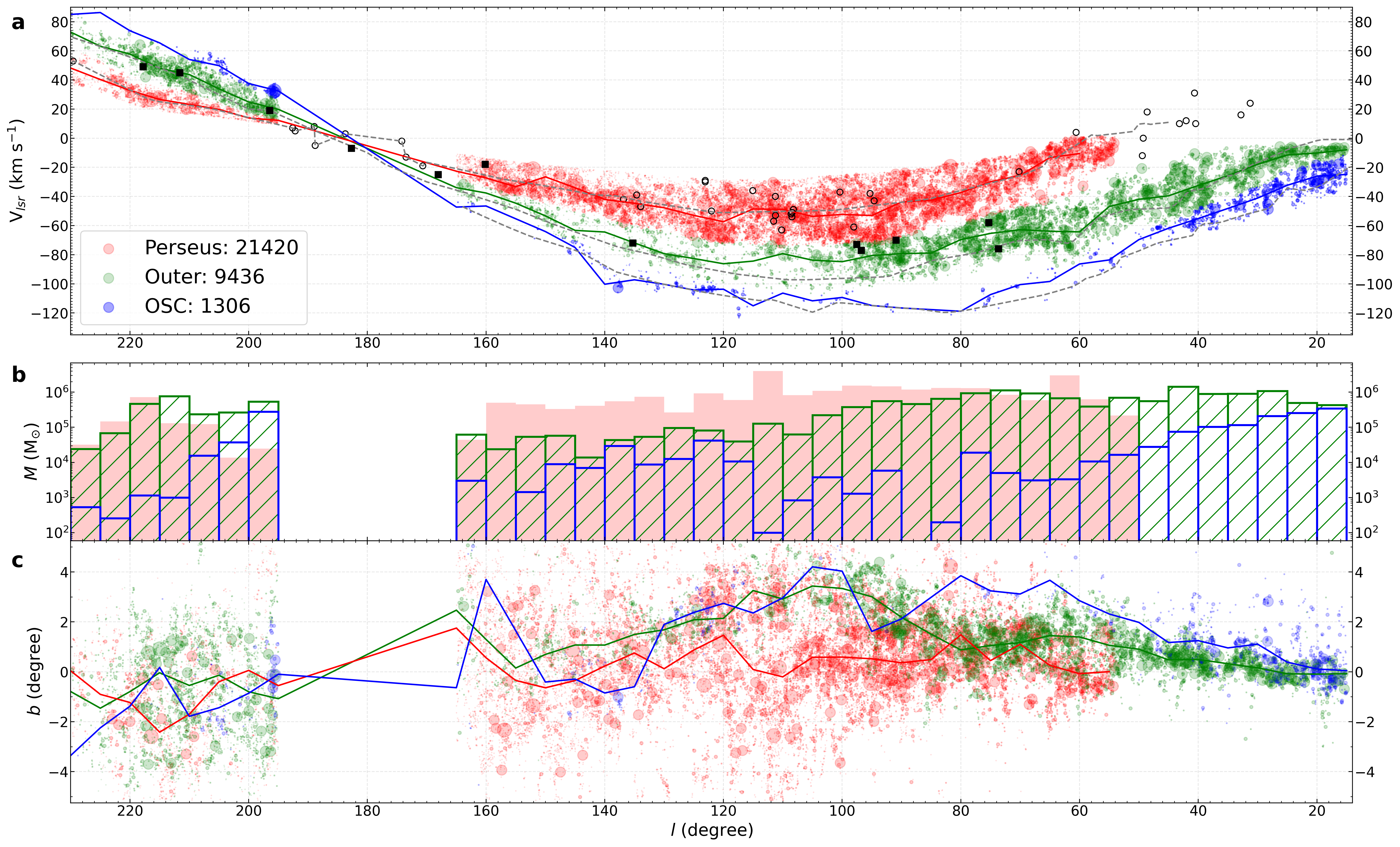}
\caption{($a$) Molecular cloud distribution in the $l$-\vlsr~ map. The total numbers of MCs in each spiral arm are indicated in the corner. The size of individual circles gives an indication of cloud mass. The smallest circle is related to molecular clouds with a mass less than 100~M$_{\odot}$, while the largest circle corresponds to those with a mass greater than 10$^6$~M$_{\odot}$.
 The grey and color-coded lines represent the spiral-arm loci from~\citet{2019ApJ...885..131R} and MCs in this study. ($b$) Histograms illustrate the mass distributions of each spiral arm along the Galactic longitude in each 5$^{\circ}$ bin. ($c$) Distribution of MCs in the Galactic coordinates ($l$, $b$). 
The HMSFRs of the Perseus and Outer arm segments are indicated by open circles and filled squares, respectively~\citep{2019ApJ...885..131R}.  
\label{fig:lbv}}
\end{figure*}
\begin{figure*}
\vspace{-1mm}
\centering
\includegraphics[angle=0,scale=0.41]{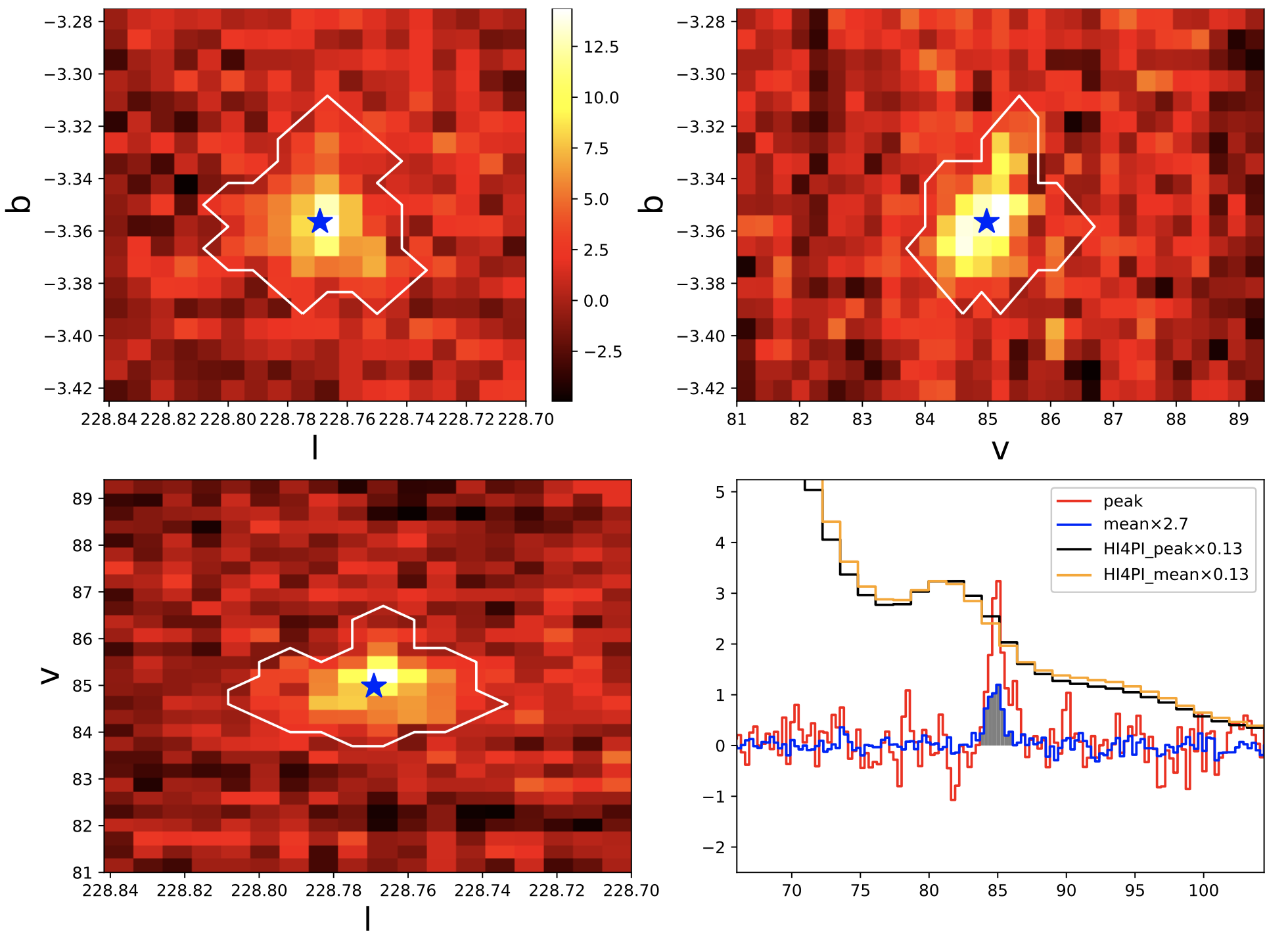}
\caption{An example of molecular clouds probably lying along the new extension of the OSC arm in the third quadrant. The blue stars indicate the center of the cloud. The white contours indicate the boundary of the cloud in the $l$-$b$-$v$ intensity maps. The CO and H{\sc i} spectra show emission at the CO emission peak and emission averaged over the white boundary.     
\label{fig:example}}
\end{figure*}

Since its initial discovery, the intense H{\sc i} 21-cm line emission originating from the atomic phase of the ISM has shown quasi-continuous features in longitude-velocity space, commonly interpreted as spiral arms~\citep[e.g.,][]{weaver1974}. H{\sc i} has provided a comprehensive view of the Galactic  disk~\citep[e.g.,][]{1957Natur.180..677K,1971A&A....10...76B,2007AJ....134.2252S,2023A&A...671A..54M}. However, despite its utility, the H{\sc i} spiral structure is still unclear or blurry in the total H{\sc i} surface density map~\citep[e.g.,][]{2006Sci...312.1773L,2009ARA&A..47...27K}. 
A critical issue is the large intrinsic velocity dispersion, which causes features to be smeared along the line of sight by confusing the velocity-distance transformation~\citep{2009ARA&A..47...27K}. 
Additionally, H{\sc i} contains both cold- and warm-phase (a few thousand K) gases and can extend even beyond the disk to the starless disk-halo interface~\citep{2023ARA&A..61...19M}, which further complicates its role as an ideal arm tracer. To address these issues, various methods have been developed, such as transforming line flux within narrow velocity intervals~\citep[e.g.,][]{1969ARA&A...7...39K,2017PASP..129i4102K}, or applying unsharp masking techniques to enhance the visibility of spiral structures~\citep{2006Sci...312.1773L}. Despite these efforts, discrepancies persist among studies even when utilizing the same observational data.

In contrast, as it traces the cold and dense component of the ISM and is closely related to star formation, CO line emission offers a more conventional arm indicator. Furthermore, owing to its small velocity dispersion, the CO emission can be well defined as a series of discrete molecular clouds (MCs), which can serve as an alternative way to construct the face-on spiral structure from the central velocity of each MC, in addition to directly constructing it from the velocity of each voxel in the (position-position-velocity) PPV cube. Therefore, CO molecular clouds are expected to delineate compelling spiral structures across the entire disk. However, limited by the combined effects of the much reduced intensity of the CO line and the sensitivity of previous CO surveys, the global CO spiral structures were determined within a much smaller Galactic range than H{\sc i}~\citep{2008ApJ...677..283P,2016PASJ...68....5N,2016ApJ...822...52R,2017ApJ...834...57M,2021A&A...655A..64M}. As a result, much of the outer Galactic molecular disk remains largely unexplored~\citep[as recently reviewed by][]{2017ASSL..434..175W}.

 The new deeper CO survey--Milky Way Imaging Scroll Painting~\citep[MWISP;][]{2019ApJS..240....9S,2021ApJS..256...32S} conducted with the PMO 13.7-m millimeter-wavelength telescope~\citep{yang2008,shan2012,sunjx2018}, 
 offers us a unique opportunity to uncover the distant large-scale molecular structures. In more recent years, some progress on large-scale molecular structures has been made based on MWISP's early data, i.e., the identification of the new segment of the distant spiral arm~\citep{2015ApJ...798L..27S,2017ApJS..230...17S}, and the apparent reveal of the Outer arm in the form of molecular gas~\citep{2016ApJ...828...59S,2016ApJS..224....7D,2023ApJS..268....1D} in the regions of $l$=[35$^{\circ}$, 45$^{\circ}$], [100$^{\circ}$, 150$^{\circ}$] and [220$^{\circ}$, 230$^{\circ}$]. 
 Currently, the MWISP survey has fully covered the Galactic coordinate ranges of $l$=[12$^{\circ}$, 230$^{\circ}$] and $b$=[$-$5.25$^{\circ}$, 5.25$^{\circ}$]. Using this dataset, we focus on the face-on spiral structures in the outer Galaxy, where velocity crowding is alleviated, and distances can be unambiguously estimated from the observed radial velocity. Although the Local arm is part of the Outer Galaxy, it is excluded from the scope of this study due to its broad distribution in Galactic latitude, which is not well covered by current MWISP data, as well as the significant uncertainty in its kinematic distance.


\section{Molecular clouds distribution in the ($l$, $b$, \vlsr) space}

We use so far the largest sample comprising 32,162 MCs, which were generated using a density-based spatial clustering of applications with the noise~\citep[DBSCAN,][]{1996kddm.conf..226E,2020ApJ...898...80Y} algorithm applied to the MWISP data. The algorithm can effectively identifies contiguous PPV voxels above a given intensity threshold and has been tested against other widely used algorithms by \citet{yan2022}. DBSCAN, like other algorithms, faces limitations in crowded regions where it may mistakenly group unrelated structures into a single cloud or, in areas with low sensitivity, split emissions that belong to the same cloud into different clouds. These are common uncertainties encountered when extracting individual clouds from PPV data cubes. 
However, this issue is largely mitigated in the lower cloud density outer Galaxy. The sample has been filtered by visual inspection of the morphology and spectra of each MC. Therefore, the MCs of this sample represent the confident ones. More details regarding the identification and properties of these MCs are presented in Sun et al. (accepted by ApJS, 2024, hereafter Paper {\sc i}). 

Figure~\ref{fig:lbv} presents the distributions of these 32,162 MCs in the Galactic coordinates. Each MC was assigned to a spiral arm by matching its Galactic longitude and LSR velocity to the nearest known spiral arm on the $l$-\vlsr~ map (see the grey-dashed lines in Fig.~\ref{fig:lbv}), which is primarily based on CO, H{\sc i}, or maser data \citep{2019ApJ...885..131R}. This method assumes that the boundaries of each spiral arm are defined by equal velocity separations from the centers of adjacent arms. The total number of MCs assigned to the three spiral arm segments is also indicated in Fig.~\ref{fig:lbv}.


The MCs along the outermost feature over a longitude range of 15$^{\circ}$-165$^{\circ}$ (Fig.~\ref{fig:lbv}a) confirm our previous hypothesis that the OSC arm might extend into the outer second quadrant~\citep{2015ApJ...798L..27S}. Most surprisingly, however, is that 
a feature at the largest positive velocity might be a new extension of the OSC arm, from ($l$, $b$, $v$)$\approx$(195$^{\circ}$, $-$1.2$^{\circ}$, 36~\kms), to (230$^{\circ}$, $-$3$^{\circ}$, 83~\kms), running roughly parallel to the locus of the Outer arm but shifted by $\sim$15~\kms~ to more positive velocities. The shifting is much larger than the typical cloud-cloud velocity dispersion of $\sim$5~km$\,$s$^{\rm -1}$ within each spiral arm, while it is comparable to the typical velocity dispersion of H{\sc i} emission. Inspection of the H{\sc i} spectra~\citep{2016A&A...594A.116H} of those clouds indeed reveals a dimmer emission peak that lies beyond the Outer arm but is not entirely separated from it (Fig.~\ref{fig:example}). These would provide natural explanations for the ignoring of the outermost feature by previous H{\sc i} studies.        

Our observed spiral-arm loci from CO are indicated by colored lines in Fig.~\ref{fig:lbv}, which were derived from the average (weighted by cloud mass) of $v_{\rm lsr}$ of MCs in each 5$^{\circ}$ bin. 
While there are slight offsets between the colored and grey lines in certain regions, they can still be regarded as consistent structures.
The histograms illustrate the molecular gas masses summed over 5$^{\circ}$ intervals, which are not evenly distributed along all spiral arms (Fig.~\ref{fig:lbv}b). For instance, the Outer arm becomes less prominent between a Galactic longitude range of 100$^{\circ}<l<$165$^{\circ}$, and the OSC arm appears dim in the range 70$^{\circ}<l<$100$^{\circ}$. Another possible reason for the scarcity of clouds in the latter region could be due to a large warp in the outskirts of our Galaxy that extends beyond our current coverage range (Fig.~\ref{fig:lbv}c). The phase II survey of MWISP with Galactic latitude coverage up to $b=\pm$10$^{\circ}$ would either confirm or reject this possibility in the near future.

\section{Face-on view of the outer Galactic molecular disk}
 \begin{figure*}
\vspace{-3mm}
\centering
\includegraphics[angle=0,scale=0.5]{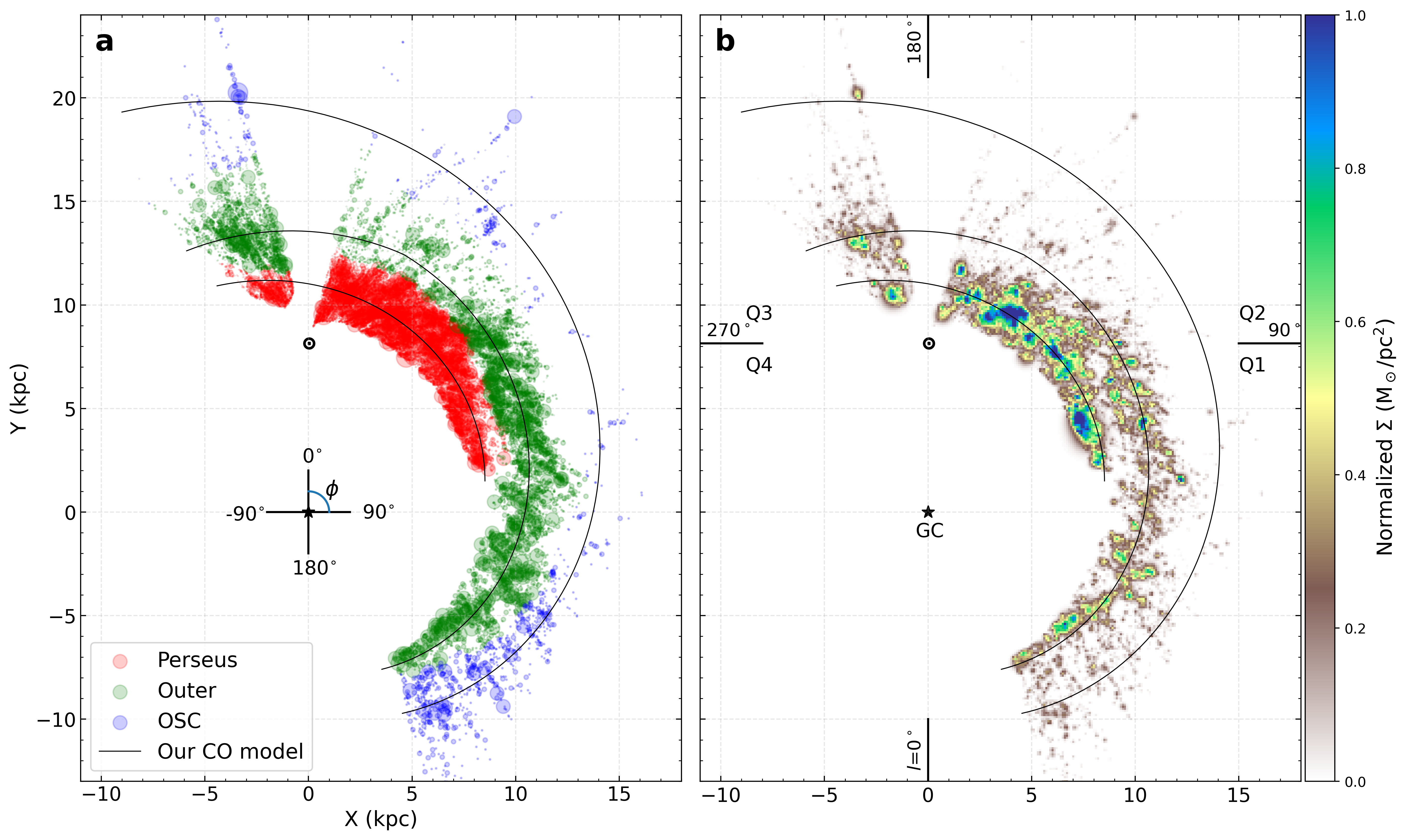}
\includegraphics[angle=0,scale=0.5]{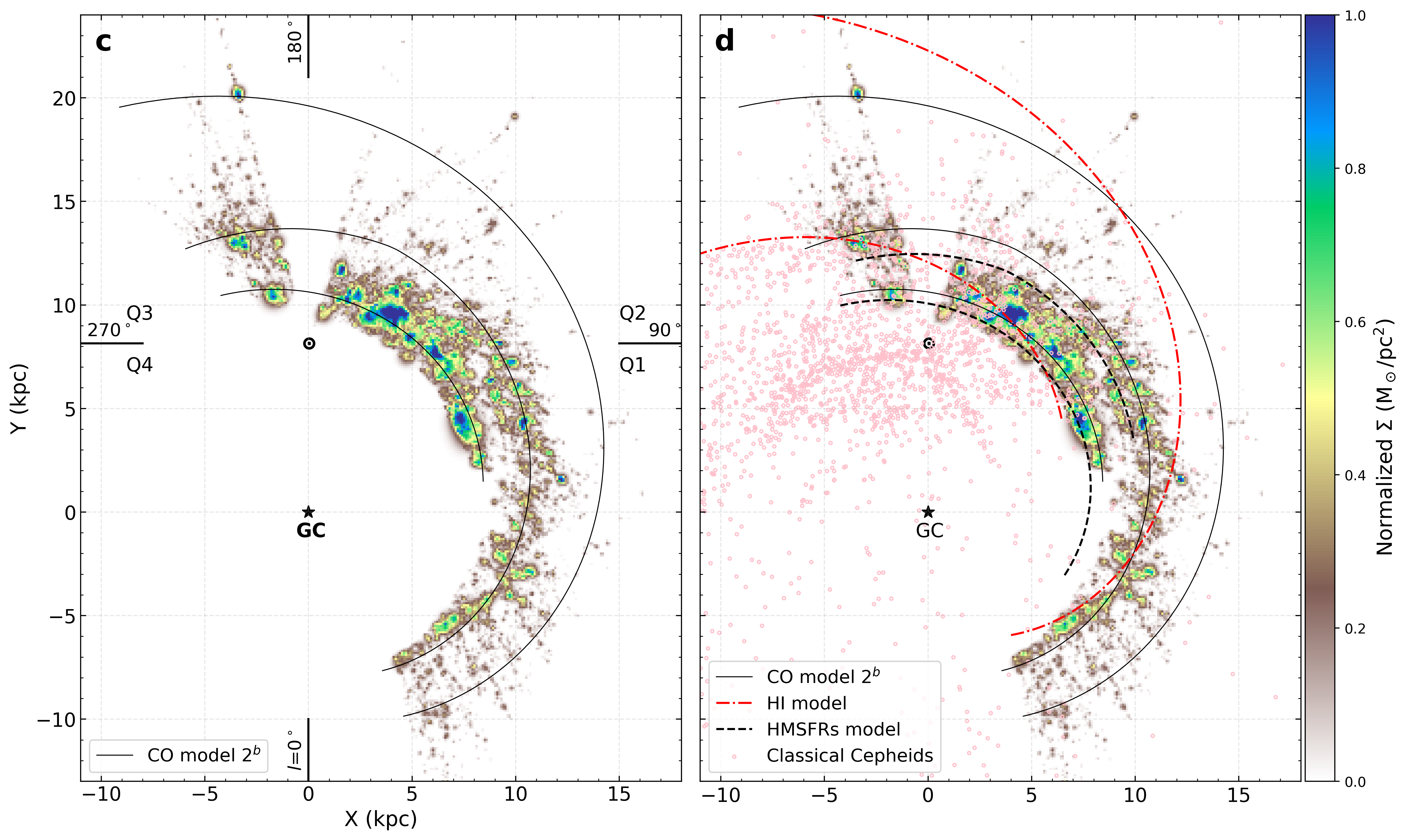}
\caption{($a$) Distribution of MCs projected into the Galactic plane. Each circle represents a single cloud with size scaled by cloud mass. The solid lines trace the log-periodic spiral fitting results to MCs weighted by mass. ($b$) Surface molecular gas density distribution map. Each MC was brightened with a Gaussian function. To highlight the main structure, the intensity is normalized to unity. Galactic quadrants are marked. ($c$) Same as panel ($b$), but with cloud masses corrected for variations in $X_{\rm CO}$ and sensitivity clip factors. ($d$) Comparison between CO and other tracers. Dashed lines are the models on the basis of selected HMSFRs in the Outer and Perseus arms~\citep{2019ApJ...885..131R}. Red dot-dashed lines represent the H{\sc i} arm models~\citep{2006Sci...312.1773L}. Pink circles indicate the classical Cepheid variable stars~\citep{2019Sci...365..478S}. 
\label{fig:arm}}
\end{figure*}
By assuming the new rotation curve and Galactic parameters from fit A5 of~\citet{2019ApJ...885..131R}, we calculated the distance for each MC and converted the heliocentric coordinates ($l$, $b$, \vlsr) to the Galactocentric cylindrical coordinates ($R$, $\phi$, $z$). The face-on distribution of MCs in the outer Galaxy is present in Fig.~\ref{fig:arm}a. 
 Within each component, the overdense region is discernible but not very clear owing to the significant overlapping of MCs. To better indicate spiral arms, we need to reduce the map along the direction perpendicular to the disk. To reproduce a top-down view of the molecular disk, we need to project each extended cloud onto a regular $x$-$y$ grid and then accumulate the emission within each cell. We assume a two-dimensional Gaussian profile for each MC~\citep[e.g.,][]{hou2009,2014A&A...569A.125H}.  
\begin{equation}\label{eq:LebsequeI}
        L(x,y)=\sum_i {\frac{W_{i}} {2\pi
    {\sigma_{i}}^2}}exp(-{\frac{(x-x_{i})^{2}+(y-y_{i})^{2}} { 2
    {\sigma_{i}}^2}}).
\end{equation}

Here, we adopt the measured mass of each cloud as the weight factor $W_i$. $x_i$ and $y_i$ are the coordinates of the $i$th MC. We choose a pixel size of 100 pc for the $x$-$y$ map, which is relatively high for calculations but still somewhat larger than the typical size of the clouds. Therefore, each MC is grided with a convolution. The convolution kernel is a Gaussian function with sigma of one-third the pixel size of the map. As a result, the standard deviation $\sigma_i$ of the profile was derived from the convolution result of the size of MC and that of the convolution kernel. Note that a map derived from the central velocity of each MC can help reduce the blurring effect caused by the variation in velocities across different parts of the cloud due to turbulence or other random motions. Hence, this method allows for better visualization of the spiral structure compared to directly converting it from the velocity of each voxel in the PPV cube.

Fig.~\ref{fig:arm}b is essentially a face-on surface density distribution map and the intensity is normalized to unity. Fig.~\ref{fig:arm}c shows the same map as Fig.~\ref{fig:arm}b, but with cloud masses corrected for variations in $X_{\rm CO}$ and sensitivity clipping factors~(Paper {\sc i}).  It is important to note that while these corrections improve the mass completeness for detected clouds, the incompleteness in detecting low-mass clouds has not yet been accounted for.  

An exciting result is that several spectacular spiral arms 
 become clearly visible even on the far side of the Milky Way, with a noticeable contrast between the arm and interarm densities. These CO spiral structures are traced out to a radius of $\sim$22 kpc in the third Galactic quadrant, which is similar to that of H{\sc i}~\citep{2006Sci...312.1773L} but extends further than previous CO studies~\citep[e.g.,][]{2008ApJ...677..283P,2016PASJ...68....5N,2016ApJ...822...52R,2017ApJ...834...57M,2021A&A...655A..64M} by about 1.5 times.  
 The sharper and more apparent CO spiral structure compared to H{\sc i} may benefit from the advantages of CO tracer with its higher angular resolution and concentration distribution along the spiral arm, as well as the method used, which minimizes the influence of smearing along the line of sight.


In Fig.~\ref{fig:arm}b and Fig.~\ref{fig:arm}c, the Perseus arm stands out as the most prominent structure along most of its length, appearing as a quasi-continuous spiral arm. The linear mass of arm segments, defined as the ratio of total mass to arm length, is also seems to be largest for the Perseus arm in the studied azimuthal direction. 
Despite being known for many years, the Outer arm is clearly visible for the first time as a molecular gas feature extending nearly 33 kpc in the outer Galaxy alone, with over 65\% of its total gas mass discovered through the MWISP survey. Beyond it, the large-scale molecular gas feature is exclusively traced by the MWISP data~(Table 2 of Paper {\sc i}). Although the dimmer OSC arm segment appears fragmented in CO emission, the OSC arm in the previously studied range still exhibits quasi-continuous H{\sc i} emission~\citep{2011ApJ...734L..24D, 2015ApJ...798L..27S, 2017ApJS..230...17S}. Next-generation surveys with even higher resolution and sensitivity can help to establish a more complete cloud catalog, thereby delineating a more continuous molecular gas distribution in this distant arm segment. 

\begin{deluxetable*}{lccccccccccccccccc}
\tabletypesize{\scriptsize}
\setlength{\tabcolsep}{0.03in}
\tablewidth{0pt}
\tablecaption{Parameters of each CO spiral arm segment derived from the MWISP data.\label{tab:sample}}
\tablehead{
 &  &  & \multicolumn{7}{c}{Constant $X_{\rm CO}$} & & \multicolumn{7}{c}{Variations in $X_{\rm CO}$ \& sensitivity clip factors}\\
 \cline{4-10} \cline{12-18} 
  &  &  & & & \multicolumn{5}{c}{Model 1} & &&& \multicolumn{5}{c}{Model 2}\\
 \cline{6-10} \cline{14-18} 
Arm & $\phi$ range & Length & Mass & M/L & $N$ & $\phi_{\rm kink}$  & $R_{\rm kink}$ & $\psi_<$ &$\psi_>$ && Mass & M/L & $N$ &  $\phi_{\rm kink}$  & $R_{\rm kink}$ & $\psi_<$ &$\psi_>$ \\
& ($\circ$)& (kpc) & (M$_{\odot}$) &  (M$_{\odot}\,$kpc$^{-1}$) && ($\circ$) & (kpc) & ($\circ$) & ($\circ$) && (M$_{\odot}$) &  (M$_{\odot}\,$kpc$^{-1}$) && ($\circ$) & (kpc) & ($\circ$) & ($\circ$) \\
(1) & (2) &(3) &(4)&(5)&(6)&(7)&(8)&(9)& (10)&&(11)&(12)&(13)&(14)&(15)&(16)&(17)}
\startdata
 Perseus$^a$ &$-$19.7$\rightarrow$$-$3.8, 1.5$\rightarrow$77.4& 16.2 & 2.4$\times$10$^7$ & 1.5$\times$10$^6$ &21,420&  30.0 & 10.1 & 9.8  & 9.8 && 4.0$\times$10$^7$ & 2.5$\times$10$^6$ &21,420& 30.0 & 10.1 & 9.7 & 9.7 \\
 Perseus$^{b}$& & & & &  601 & 30.0 & 9.7 & 9.0 & 9.0 && & &  499 & 30.0 & 9.8 & 8.8 & 8.8 \\
Outer$^a$   & $-$26.8$\rightarrow$$-$4.7, 6.2$\rightarrow$150.4 & 32.9 & 1.6$\times$10$^7$ &  4.9$\times$10$^5$&9,436&  20.3 & 13.3 & 3.5  &11.1 && 3.3$\times$10$^7$ &1.0$\times$10$^6$ &9,436& 17.1 & 13.5 & 3.2 & 11.1 \\
Outer$^b$ & & & & &  1,826 & 20.6 & 13.2 & 3.6 & 11.0 && & &  1,795 & 17.8 & 13.4 & 3.5 & 10.9 \\
OSC$^a$     & $-$27.3$\rightarrow$$-$8.3, 9.6$\rightarrow$155.8 & 43.4 & 1.7$\times$10$^6$ &  3.9$\times$10$^4$&1,306 & 47.0 & 16.2 & 12.3 & 12.3 && 4.8$\times$10$^6$ &1.1$\times$10$^5$ &1,306& 47.0 & 16.3 & 12.2 & 12.2 \\
OSC$^b$ & & & & &  291 & 47.0 & 16.2 & 12.5 & 12.5 && & &  367 & 47.0 & 16.4 & 12.3 & 12.3 \\
\enddata
\tablecomments{Column (1): arm segment. In case $``a"$, we rely on an unselected cloud sample for the spiral arm fitting. In case $``b"$, those MCs with mass below the mean value of clouds' mass of the Perseus arm are excluded, specifically, with $M$<1.1$\times$10$^{\rm 3}$ M$_{\odot}$ are excluded for model 1 fit, and $M<$2.8$\times$10$^{\rm 3}$ M$_{\odot}$ are excluded for model 2 fit. The Perseus arm clouds in the second Galactic quadrant are further excluded due to their significant large uncertainty of kinematic distance~~\citep[e.g.,][]{2006Sci...311...54X,2022ApJ...925..201P}. Additionally, MCs of the Perseus arm in the $\phi$>72$^{\circ}$ are also excluded for the log-periodic spiral fitting considering the potential incompleteness of cloud sample. Column (2): $\phi$ range of the arm segment traced by MCs. Columns (3)-(6): total length and total mass of the arm segment traced by the MWISP data, ratio between the total mass and total length of each arm segment. Column (6): the number of MCs used to arm fitting. Columns (7)-(10): parameters from log-periodic spiral fitting that under the consumption of constant CO-to-H$_2$ conversion factor. $\psi_<$ and $\psi_>$ are pitch angles for azimuths <$\phi_{\rm kink}$ and $\geqslant\phi_{\rm kink}$, respectively. The case $\psi_<=\psi_>$ indicates a constant pitch angle obtained. Columns (11)-(17): same as Cols. (4)-(10) but incorporate variations of $X_{\rm CO}$ and sensitivity clip factors.}
\end{deluxetable*}

\section{CO Spiral Arm Model\label{sec:model}}

Following \citet{2019ApJ...885..131R}, we attempt to fit the distribution of the MCs using a Markov chain Monte Carlo (MCMC) approach. The adopted log-periodic spiral model form can be expressed as, ln$(R/R_{\rm kink})=-(\phi-\phi_{\rm kink})$tan$\psi$, where $R$ is the Galactocentric radius, $\phi$ is the Galactocentric azimuth, and $\psi$ is the pitch angle. 
Both uniform and abrupt changes in pitch angle on either side of the ``kink'' (with a radius of $R_{\rm kink}$ and an azimuth $\phi_{\rm kink}$) are tested. The data are variance-weighted by the cloud mass. The cloud mass derived by constant $X_{\rm CO}$ (termed as ``model 1" in Table~1) and varied $X_{\rm CO}$ and sensitivity clipping factors (``model 2") are both considered. It seems that varying $X_{\rm CO}$ and flux clip factors only marginally affects the obtained CO spiral arm model.

We fitted the model using both the full sample (referred to as ``a" in Table 1) and a subsample (termed ``b" in Table~\ref{tab:sample}), which excludes clouds with masses below the mean mass of MCs in the Perseus arm, as well as MCs that may have significantly large kinematic distances and regions with potentially lower completeness (see notes in Table 1). Since more massive clouds are more likely to be located within the arm regions, model ``b" is expected to provide a more accurate fit. The resulting log-periodic spiral models for ``b" are shown as solid lines in Fig.~\ref{fig:arm}. The Outer arm can be better fitted by a model with varying pitch angles, i.e., $\psi$=10.9$^{\circ}$ for the segment with $\phi>$17.8$^{\circ}$ and $\psi$=3.5$^{\circ}$ for the other segment. While a constant pitch angle of 12.3$^{\circ}$ fits the MCs in the current OSC arm segment, even including those in the third quadrant. This further provides support to our suggestion of the far extension of the OSC arm into the third quadrant. 
For the Perseus arm, excluding the second Galactic quadrant—where there is significant uncertainty in kinematic distances—allowed for a better fit of gas emissions in the first and third quadrants. The resulting pitch angle for this model is 8.8$^{\circ}$(Table~\ref{tab:sample}).


The geometries of the spiral arms potentially containing clues of their origin have also been extensively modeled by various tracers and methods~\citep[e.g.,][]{2008gady.book.....B,2014PASA...31...35D}. \citet{2006Sci...312.1773L} yielded pitch angles ranging from  21$^{\circ}$ to 25$^{\circ}$ by fitting the overdense peak of H{\sc i}, which is partially confirmed by the overdense maps of young upper main sequence stars, classical Cepheids, and young clusters~\citep[e.g.,][]{2021A&A...651A.104P,2023A&A...674A..37G,drimmel2024}. Other tracers like high mass star-forming regions (HMSFRs) or H{\sc ii} regions indicated a much shallower pitch angle~\citep[e.g.,][]{1976A&A....49...57G,1993ApJ...411..674T,2007A&A...470..161R,2019ApJ...885..131R,2023ApJ...947...54X}. Although the quasi-stationary density wave theory predicts spatial displacements between different arm components~\citep[e.g.,][]{1964ApJ...140..646L,1969ApJ...158..123R,2015MNRAS.454..626H}, it may not fully account for the large discrepancy reported between different tracers. Our CO models do not favor high-pitch-angle arms (Fig.~\ref{fig:arm}d). It is important to note that material is not always evenly distributed along the spiral arm, which may make the dim segment even less evident in the overdense $x$-$y$ map. Therefore, 
caution should be taken when fitting the arm model solely by connecting the overdense arm segments in the $x$-$y$ map, as this may lead to an irrational connection between the segments due to uneven material distribution along the spiral arm.

Compared with the more precise parallax-based model using HMSFRs \citep{2019ApJ...885..131R}, we find that the two known CO spiral arms align well in shape with those traced by HMSFRs, though with slight positional offsets: approximately 0.5 kpc for the Perseus arm and about 1 kpc for the Outer arm (see Fig.~\ref{fig:arm}d). These offsets may result from differences in arm tracers and uncertainties in kinematic distances. For instance, an LSR velocity uncertainty of around 5~\kms~could lead to an average kinematic distance uncertainty of about 0.47 kpc for the Perseus arm and 0.95 kpc for the Outer arm. Given these potential sources of variation, we do not address these offsets further in our analysis. 
 We should also note that a smaller distance for the Perseus arm and Outer arm would only make the arms stand out more clearly in our analysis. Therefore, while the uncertainty of the kinematic distance may affect some details, it will not significantly alter the overall picture of the CO spiral patterns. 
 
Figure~\ref{fig:arm}d also shows a comparison with the spatial distribution of the classical Cepheids~\citep{2019Sci...365..478S}, which are thought to be more numerous than young clusters~\citep{2021A&A...651A.104P,minniti2021,2023A&A...674A..37G}. However, the Cepheids are still not numerous enough to well define the spiral structure, particularly on the far side of the Milky Way. In summary, while significant progress is being made for each tracer, there is still a long way to go before uncovering and fully understanding the displacements between various tracers. 

 \section{Galactic outskirts}
 \begin{figure*}[h!]
\centering
\includegraphics[angle=0,scale=0.7]{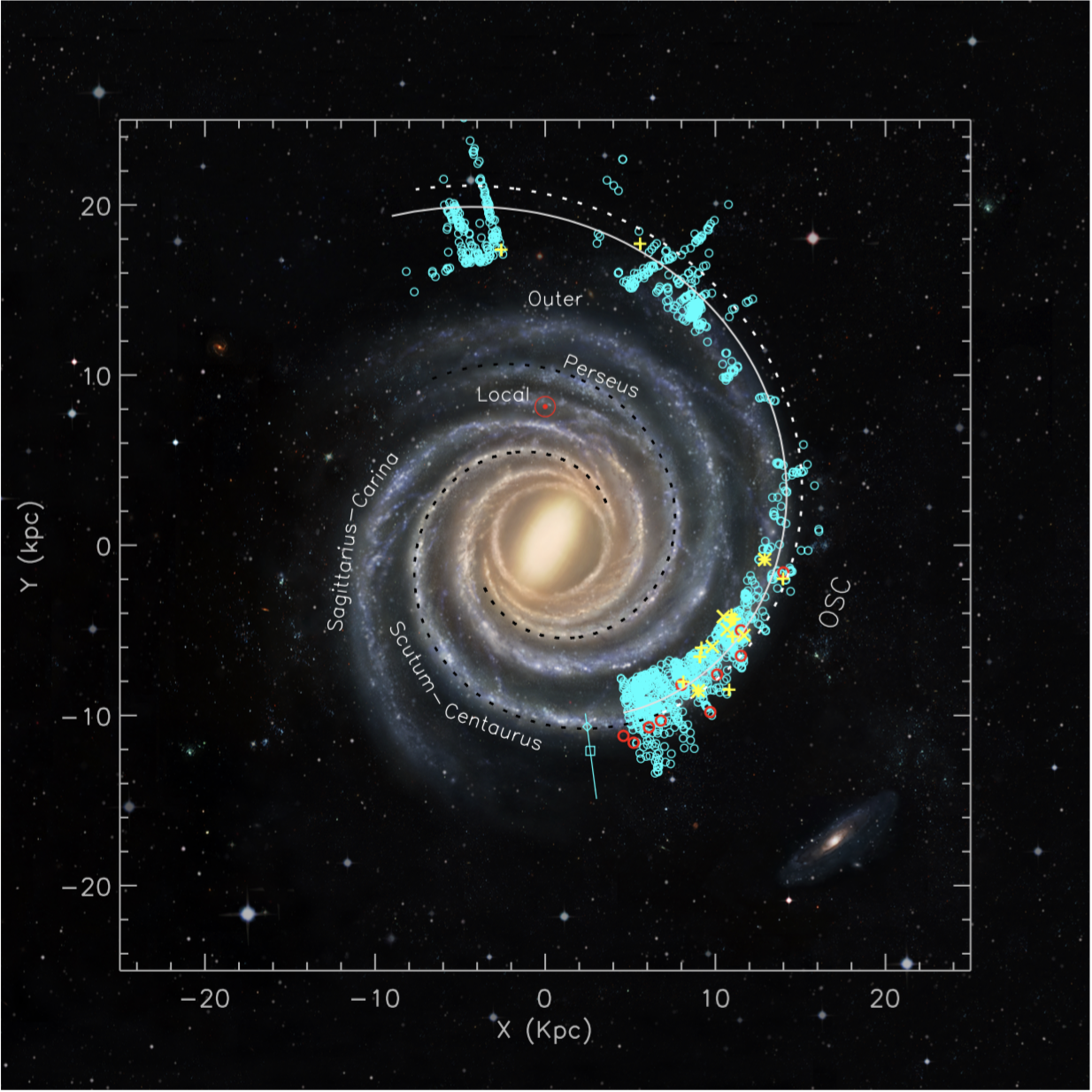}
\caption{Molecular clouds and HMSFRs at the Galactic outskirts. The available distant MCs are superposed on the new view of the Milky Way based on 200 trigonometric parallaxes for masers in HMSFRs~(credit: Xing-Wu Zheng \& Mark Reid BeSSeL/NJU/CfA). The 1306 MCs from this study and 10 MCs from the literature~\citep{2011ApJ...734L..24D} are indicated by cyan and red circles, respectively. 
The two symmetrical spirals and their extensions--namely the Scutum-Centaurus and Perseus arms, are indicated by two logarithmic spirals with a pitch angle of 12$^{\circ}$ averaged over various tracers~\citep{2008AJ....135.1301V}. The solid logarithmic spiral indicates the fit for the OSC arm (CO model 2$^b$).
The cyan square denotes the HMSFR of the OSC arm with trigonometric parallax measurement~\citep{2017Sci...358..227S}. The diamond marks the location of the HMSFR calculated by the kinematic method. The yellow pluses and crosses indicate the distant H{\sc ii} regions~\citep{Anderson2012,Anderson2015,Armentrout2017} and star-forming masers~\citep{Armentrout2017,sun2018}, respectively. 
\label{fig:osc}}
\end{figure*}

The 1306 distant MCs not only well delineate the outermost structures of our Galaxy, they serving as so far the largest cloud sample possessed low density and low metallicity, are also valuable targets for a broad range of research~\citep[e.g.,][]{2017ASSL..434..175W,2022MNRAS.510.4436M,2022ApJ...941....3K,braine2023,fontani2024}. 
This sample, together with ten MCs~\citep{2011ApJ...734L..24D}, 14 H{\sc ii} regions~\citep{Anderson2012,Anderson2015,Armentrout2017}, and seven star-forming masers~\citep{2017Sci...358..227S,sun2018,Armentrout2017} in the OSC arm from literature, is superposed on a recent conception of the Milky Way provided by the Bar and Spiral Structure Legacy (BeSSeL) Survey (see Fig.~\ref{fig:osc}). Note that the H{\sc ii} regions include 12 listed in Table 6 of \citet{Armentrout2017}, along with G149.746$-$0.199 and G195.648$-$0.106. The seven star-forming masers are also summarized in Table 3 of \citet{vallee2020}. The log-periodic lines with a constant pitch angle of 12$^{\circ}$ indicate the two approximately symmetrical spirals in the inner Galaxy (black line) and their continuation in the outer Galaxy (white line)--namely the Scutum-Centaurus and Perseus arms. 
These arms have long been assumed to be density wave arms starting at opposite ends of the Galactic bar~\citep[e.g.,][]{2000A&A...358L..13D,2008ApJ...683L.143D,2009IAUS..254..319B,2009PASP..121..213C}, although convincing evidence has remained elusive~\citep[e.g.,][]{2011ApJ...734L..24D}. 

Evidence for such models involves tracing spiral arms on the far side of the Galaxy due to their characteristic of long and symmetric shape~\citep{2011ApJ...734L..24D}. It was not until 2011, however, that the mirror-symmetrical counterpart of the Perseus arm was discovered, depicted by the ten distant MCs along the extension of the Scutum-Centaurus arm in the outer first quadrant~\citep[e.g.,][]{2011ApJ...734L..24D}.
Adding a large number of MCs along the extension of the Scutum-Centaurus arm makes the whole arm even longer, extending over a length of $\sim$80 kpc and wrapping a full 360$^{\circ}$ around the Galaxy, though in a ``looser'' or less regular manner. Our findings provide some robust evidence supporting the view of symmetry density wave arms. Future high-quality data from the southern sky are called for to reveal the expected extension of the Perseus arm as the symmetric counterpart of the present new segment. 

 \begin{acknowledgments}
 This research made use of the data from the Milky Way Imaging Scroll Painting (MWISP) project, which is a multi-line survey in $^{12}$CO/$^{13}$CO/C$^{18}$O along the northern galactic plane with the PMO-$13.7\rm~m$ telescope. We are grateful to all the members of the MWISP working group, particularly the staff members at the PMO-$13.7\rm~m$ telescope, for their long-term support. We sincerely thank the anonymous referee for the feedback and comments that helped to improve the paper. MWISP was sponsored by National Key R\&D Program of China with grants 2023YFA1608000 \& 2017YFA0402701 and by CAS Key Research Program of Frontier Sciences with grant QYZDJ-SSW-SLH047. Y.S. is supported by the Youth Innovation Promotion Association, CAS (2022085) and the “Light of West China” Program (No. xbzg-zdsys-202212). J.Y. is supported by the National Natural Science Foundation of China through grant 12041305.
\end{acknowledgments}
\vspace{5mm}
\facility{PMO-$13.7\rm~m$ telescope}






\bibliographystyle{aasjournal}
\bibliography{main} 

\end{document}